\documentclass[prl,twocolumn,amsmath,amssymb,showpacs]{revtex4}

\usepackage{amssymb}
\usepackage{graphicx}
\usepackage{amsmath}

\begin{document}

\title{Temperature Dependence of the Intrinsic Anomalous Hall Effect in Nickel}

\author{Li Ye}

\affiliation{State Key Laboratory of Surface Physics and Department
of Physics, Fudan University, Shanghai 200433, China}

\author{Yuan Tian}

\affiliation{State Key Laboratory of Surface Physics and Department
of Physics, Fudan University, Shanghai 200433, China}

\author{Di Xiao}

\affiliation{Materials Science and Technology Division, Oak Ridge
  National Laboratory, Oak Ridge, TN 37831, USA}

\author{Xiaofeng Jin}

\email[Corresponding author. ]{xfjin@fudan.edu.cn}
\affiliation{State Key Laboratory of Surface Physics and Department
of Physics, Fudan University, Shanghai 200433, China}

\date{\today}

\begin{abstract}
We investigate the unusual temperature dependence of the anomalous
Hall effect in Ni. By varying the thickness of the MBE-grown Ni
films, the longitudinal resistivity is uniquely tuned without
resorting to doping impurities; consequently, the intrinsic and
extrinsic contributions are cleanly separated out. In stark contrast
to other ferromagnets such as Fe, the intrinsic contribution in Ni
is found to be strongly temperature dependent with a value of 1100
$(\Omega\cdot\text{cm})^{-1}$ at low temperatures and 500
$(\Omega\cdot\text{cm})^{-1}$ at high temperatures. This pronounced
temperature dependence, a cause of long-standing confusion
concerning the physical origin of the AHE, is likely due to the
small energy level splitting caused by the spin orbit coupling close
to the Fermi surface. Our result helps pave the way for the general
claim of the Berry-phase interpretation for the AHE.

\end{abstract}


\maketitle

Recent years have seen a surge of renewed interest in the anomalous
Hall effect (AHE) in ferromagnets, largely driven by its close
relation to various spintronic applications\cite{1}.  It is now
firmly established that there are several competing mechanisms
contributing to the AHE.  One is the extrinsic mechanism based on
the modified impurity scattering in the presence of the spin-orbit
coupling (SOC), i.e., the skew scattering and the side jump
mechanism\cite{2,3}. The other stems from the anomalous velocity of
the Bloch electrons induced by the SOC, originally proposed by
Karplus and Luttinger\cite{4}. This contribution can be interpreted
in terms of the Berry curvature of occupied Bloch states, and is of
intrinsic nature\cite{5,6,7}.  Both recent experiments and
theoretical calculations seem to suggest that the intrinsic
contribution is dominant in moderately conducting samples of
itinerant ferromagnets\cite{8,9,10,11}.

Despite the latest success of the Berry-phase interpretation, it
comes as a surprise that the physical origin of the AHE in Ni, one
of the simplest yet most important itinerant ferromagnets, is still
unresolved.  The main problem lies in the complicated temperature
dependence of the AHE in this material, which prevents a clear
identification of different contributions.  Smit first reported a
power law scaling between the anomalous Hall resistivity
$\rho_\text{AH}$ and the longitudinal resistivity $\rho_{xx}$ with
$\rho_\text{AH}\propto\rho_{xx}^{1.4}$ in Ni\cite{2}, which does not
fit either the extrinsic or the intrinsic scenario.  This situation
is further complicated by the strong temperature dependence of the
scaling observed by Lavine\cite{12}, with
$\rho_\text{AH}\propto\rho_{xx}^{1.10}$ at low temperatures below
150~K, $\rho_\text{AH}\propto\rho_{xx}^{1.97}$ at high temperatures
around 300~K, and $\rho_\text{AH}\propto\rho_{xx}^{1.70}$ in
between.  At even lower temperature (4.2~K), Fert~\textit{et al.}
found a linear relation of $\rho_\text{AH}\propto\rho_{xx}$ by
varying the impurity concentration in their high quality Ni single
crystal samples ($\rho_{xx} < 1$ $\mathrm{\mu\Omega\cdot cm}$),
which is a clear indication of the extrinsic skew scattering
mechanism\cite{13}.  Theoretical investigations have not provided
much insight either.  Large discrepancy was found between
first-principles calculations of the intrinsic AHE and
experiments\cite{14}. This is in sharp contrast to other itinerant
ferromagnets such as Fe and Co, where agreement within better than
30\% has been found and the dominance of the intrinsic mechanism is
clearly established\cite{8,11,23}.

In this Letter we examine the temperature dependence of the AHE in
Ni thin films with varying thickness. This approach allows us to
tune the resistivity $\rho_{xx}$ without modifying the impurity
concentration.  We are able to extract, for the first time, the
intrinsic anomalous Hall conductivity in Ni, and demonstrate
quantitatively its dominance. Surprisingly, the intrinsic AHE is
strongly temperature dependent over a large range (5 to 150 K) in
which the magnetization stays almost the same. This is different
from the previously studied cases where the temperature dependence
is explained by changes in the magnetization\cite{24,25}. We
attribute the strong temperature dependence to the complex Fermi
surface of Ni in the presence of the SOC.  Our result not only
clears up the long-standing puzzle of the physical origin of the AHE
in Ni,  but also paves the way for the general claim of the
Berry-phase interpretation for the AHE.

\begin{figure}
\center
\includegraphics[width=0.42\textwidth]{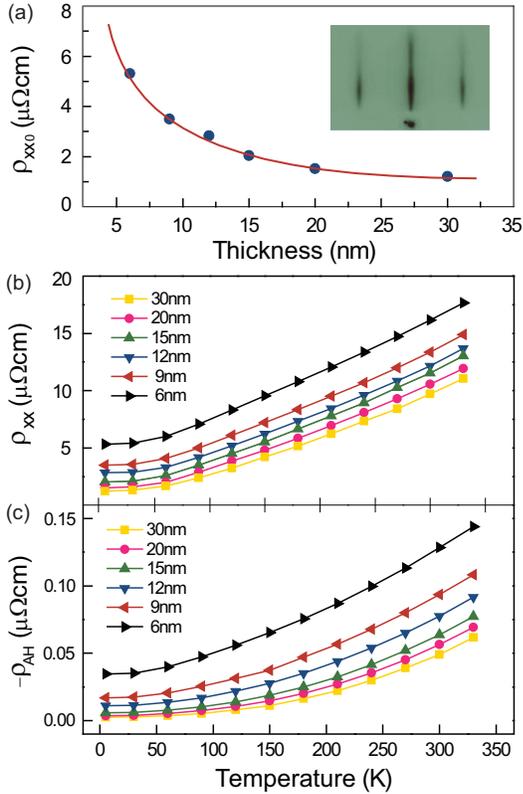}
\caption{(a) Thickness dependence of $\rho_{xx0}$ with red guideline
for eyes. Inset: RHEED pattern for 30nm thick Ni film, with incident
electron along MgO [100]. (b), (c) Temperature dependence of
$\rho_{xx}$ and $\rho_\text{AH}$ for various film thicknesses.}
\end{figure}

Ni films with thickness from 6 to 30~nm were epitaxially grown on
MgO(001) by molecular beam epitaxy with its orientation relative to
the substrate, Ni[001]$\parallel$MgO[001] and
Ni[100]$\parallel$MgO[100]; the detailed experimental setup was
described elsewhere\cite{15}.  Clean and ordered MgO(001) surface
was first prepared by annealing at 1100~K in UHV, on which the
epitaxial growth of Ni at 300~K was then followed.  Using an in-situ
mask, we have prepared films with several different thicknesses on
the same substrate.  They were further annealed at 600~K for 1 hour
to acquire better crystal quality and surface morphology\cite{16}. A
representative reflection high energy electron diffraction (RHEED)
pattern after annealing is shown in the inset of Fig.~1(a). The
improved sample quality after the annealing was also revealed from
the significant decreasing ($>50\%$) of the corresponding sample
residual resistivity $\rho_{xx0}$.  Meanwhile, the magnetization
monitored by the magneto-optic Kerr effect (MOKE) remained almost
the same, which indicated that the interface diffusion between Ni
and MgO was negligible during the annealing. In order to prevent
oxidation in the ambient air during the transport measurement, a
capping layer of 5~nm MgO was further deposited on top of Ni before
each sample was taken out from the UHV chamber. The films were then
patterned into the form of a standard Hall bar along [100] with the
magnetic field along the [001] direction, and the anomalous Hall
resistivity $\rho_\text{AH}$ and the longitudinal resistivity
$\rho_{xx}$ were simultaneously measured with the physical property
measurement system (PPMS-9T).

Figure~1(a) shows the residual resistivity $\rho_{xx0}$ of Ni
(measured at 5~K) as a function of film thickness ranging from 6 to
30~nm, where a factor of five-fold decrease is observed. This is due
to the finite size effect in electrical resistivity of thin metallic
films induced by the geometrical limitation of the bulk (or
background) mean free path of conduction electrons\cite{17}. The
current selection of film thicknesses allows us to change the
impurity scattering with little alteration of the bulk electronic
structure \cite{18,19}.

Figure~1(b) and (c) show the temperature dependence of both
$\rho_{xx}$ and $\rho_\text{AH}$ for different film thicknesses,
respectively.  The negative sign of $\rho_\text{AH}$ reflects the
fact that the chirality of the AHE in Ni is opposite to that of Fe
\cite{11}. We plot $\rho_\text{AH}$ as a function of $\rho_{xx}$ for
the thickest 30 nm Ni film in Fig. 2(a).  The
$\rho_\text{AH}=f(\rho_{xx})$ curve agrees well with previous
observations for the temperature-dependent scaling of the AHE in
bulk Ni \cite{12}.  As mentioned earlier, it is exactly this
complicated scaling that makes the intrinsic origin of the AHE in Ni
rather elusive.

We now try to separate the different contributions to the AHE by
applying the scaling law proposed in our recent work, in which we
have shown that the extrinsic contribution should be scaled against
the residual resistivity $\rho_{xx0}$ instead of the total
resistivity $\rho_{xx}$.  This leads to the following scaling
relation\cite{11}:
\begin{equation}
\sigma_\text{AH}=-(\alpha\sigma^{-1}_{xx0}+\beta\sigma^{-2}_{xx0})\sigma^{2}_{xx}-b \;,
\end{equation}
where $\alpha$, $\beta$, and $b$ are constants to be determined from
the fitting, and $\sigma_\text{AH}$, $\sigma_{xx0}$, and
$\sigma_{xx}$ are the anomalous Hall conductivity, residual and
total longitudinal conductivity, respectively.  Here, the first and
last terms represent the extrinsic skew scattering and the intrinsic
Berry phase contributions respectively, while the $\beta$ term is
clearly extrinsic and was previously ascribed to the side-jump
contribution. Figure~2(b) shows $\sigma_\text{AH}$ as a function of
$\sigma_{xx}^{2}(T)$ from 5 to 330 K for different film thicknesses.
It is observed, as anticipated, that when $\sigma_{xx}^{2}(T)$ goes
to zero, the anomalous Hall conductivity $\sigma_\text{AH}$ for
various film thicknesses with different residual resistivity
converges to a universal value $500\mathrm{\Omega^{-1}cm^{-1}}$,
reflecting unambiguously the intrinsic nature of the AHE at the high
temperature limit.  However, for a given thickness, significant
deviation from a linear scaling between $\sigma_\text{AH}$ and
$\sigma_{xx}^2(T)$ are clearly noticeable, which suggests that the
scaling in Eq.~(1) does not apply directly to the AHE in Ni, if
$\alpha$, $\beta$ and $b$ were to be fixed at constant values.

\begin{figure}
\center
\includegraphics[width=0.425\textwidth]{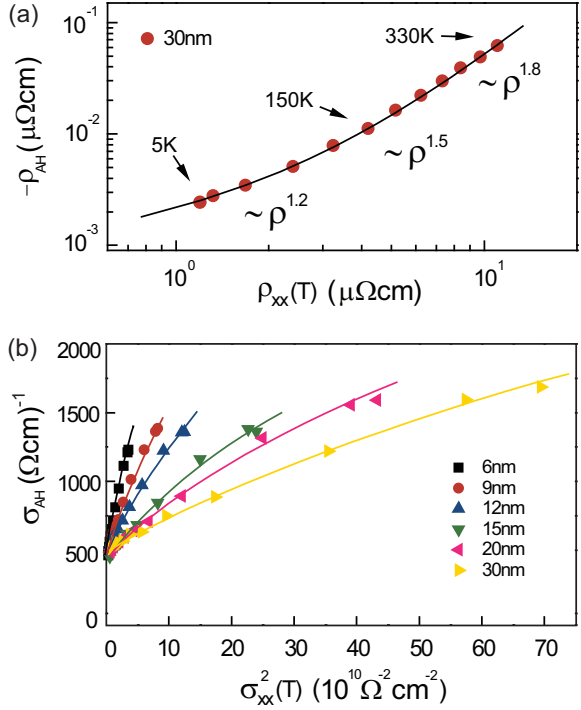}
\caption{(a) $\rho_\text{AH}$ v.s. $\rho_{xx}(T)$ plot in
logarithmic scale for 30nm-Ni showing temperature dependent power
law. (b)$\sigma_\text{AH}$ v.s. $\sigma^{2}_{xx}(T)$ plot for
various film thicknesses.}
\end{figure}

To remedy this situation, we propose
a generic scaling
\begin{equation}
\sigma_\text{AH}=-(\alpha\sigma^{-1}_{xx0}+\beta\sigma^{-2}_{xx0})\sigma^{2}_{xx}-b(T)
\;,
\end{equation}
or equivalently
\begin{equation}
\rho_\text{AH}=(\alpha\rho_{xx0}+\beta\rho^{2}_{xx0})+b(T)\rho^{2}_{xx} \;,
\end{equation}
where $\alpha$ and $\beta$ are still constants, but $b(T)$, which
represents the intrinsic Berry-phase contribution, is now a function
of temperature.  Recall that the magneto-crystalline anisotropy of
Ni is strongly temperature dependent between 5 and 330 K\cite{20}.
It is thus reasonable to anticipate that the Berry-phase
contribution to the AHE, which originates from the SOC in the band
structure, is also temperature-dependent.  Therefore, instead of
analyzing the scaling $\rho_\text{AH}=f(\rho_{xx}(T))$ with varying
temperature for each fixed film thickness, we should consider the
same scaling but with varying film thickness for each fixed
temperature $\rho_\text{AH}=f(\rho_{xx}(d))$.

We first consider the low-temperature limit.  At 5 K, the phonon
scattering is negligible and the scaling relation in Eq.~(3) reduces
to $\rho_\text{AH0}=\alpha\rho_{xx0}+(\beta+b_{0})\rho^{2}_{xx0}$,
where $b_{0}=b$ ($T=5 K$) is a constant for different film
thickness. Figure~3(a) shows the $\rho_\text{AH0}/\rho_{xx0}(d)$
v.s. $\rho_{xx0}(d)$ plot, using the set of 5 K data in Fig.~1(b)
and (c). From the nice linear fitting by the black line as shown in
the figure, we obtain the skew scattering constant
$\alpha=-7\times10^{-4}$ directly from the intercept, which is
comparable with the value (on the order of $10^{-3}$ depending on
impurity type) previously obtained by Fert~\textit{et al}\cite{13}.

Now the scaling can be recast into
$\rho_\text{AH}-\alpha\rho_{xx0}=\beta\rho_{xx0}^{2}+b(T)\rho_{xx}^{2}$
, and Fig. 3(b) shows the new set of plots
$\rho_\text{AH}(d)-\alpha\rho_{xx0}(d)$ v.s. $\rho^{2}_{xx}(d)$ with
varying film thickness for each fixed temperature ranging from 5 to
330 K.  It is clear that at each given temperature all the
experimental data can indeed be well described by the generic
scaling, as they are all nicely fitted by the straight lines in the
whole temperature range. In addition, it is found that
$\beta\approx0$ in the specific MgO/Ni/MgO system, while it can be
tuned to nonzero for different interfaces (not shown here).

\begin{figure}
\center
\includegraphics[width=0.4\textwidth]{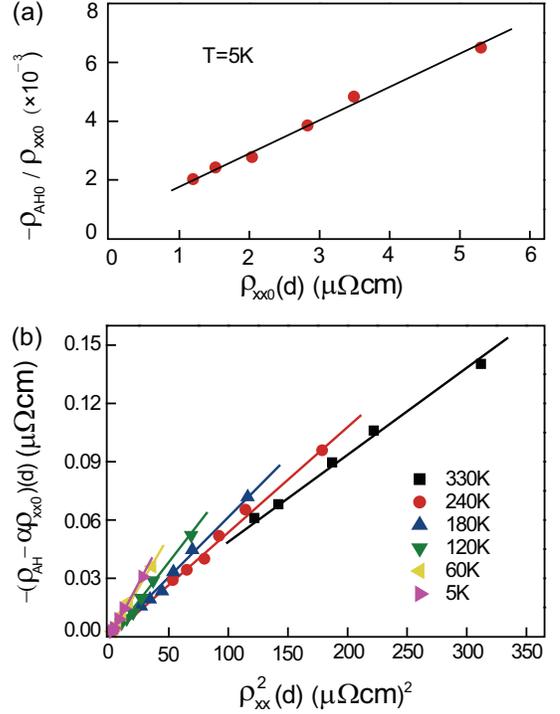}
\caption{(a)$\rho_{AH0}/\rho_{xx0}(d)$ v.s. $\rho_{xx0}(d)$ plot.
(b)$\rho_\text{AH}(d)-\alpha\rho_{xx0}(d)$ v.s. $\rho^{2}_{xx}(d)$
plot at various temperatures.}
\end{figure}

The different slopes in Fig.~3(b) at each given temperature give the
intrinsic temperature-dependent anomalous Hall conductivity
$\sigma_\text{int}(T)=-b(T)$ in Ni, which are shown in Fig.~4,
marked by solid (green) circle.  $\sigma_\text{int}(T)$ is about
1100 $(\Omega\cdot\text{cm})^{-1}$ at low temperature, and decreases
to about 500 $(\Omega\cdot\text{cm})^{-1}$ when the temperature is
above 300 K.  It becomes transparent now that it is this temperature
dependent intrinsic term that has caused all the earlier
complications and confusions in understanding the AHE in Ni. It
should be pointed out that this term cannot be singled out from the
AHE measurements on a single Ni sample.  Instead, it is only
possible on a series of samples with tunable residual resistivity
$\rho_{xx0}$ while keeping their electronic band structures
basically the same, otherwise the corresponding Berry curvatures
would be different from each other. This argument also explains why
the previous approaches by adding impurities in Ni have all failed
to unveil the intrinsic origin for the AHE in Ni.

To better understand the microscopic mechanisms of the AHE in bulk
Ni, we also show in Fig.~4 the raw data (red square dots) of the
experimentally measured anomalous Hall conductivity
$\sigma^{30nm}_\text{AH}$ in the bulk-like 30 nm Ni film. Using the
aforementioned result of $\alpha=-7\times10^{-4}$ we can plot the
temperature dependent skew scattering conductivity
$\sigma^{30nm}_\text{sk}=-\alpha\sigma_{xx0}^{-1}\sigma_{xx}^{2}(T)$
for this bulk-like 30 nm Ni film, shown as blue triangles in
Fig.~4(a). Now subtracting $\sigma^{30nm}_\text{sk}$ from
$\sigma^{30nm}_\text{AH}$, we obtain the intrinsic anomalous Hall
conductivity in this 30 nm Ni film, as given by the open circles in
Fig. 4(a). It should be emphasized that $\sigma_\text{int}(T)$ and
$\alpha=-7\times10^{-4}$ are obtained from a series of samples with
different film thicknesses (not necessary to include the data from
the 30 nm Ni film) while $\sigma^{30nm}_\text{AH} -
\sigma^{30nm}_\text{sk}$ is for a single film thickness.  The
excellent agreement between $\sigma_\text{int}(T)$ and
$\sigma_\text{int}^{30nm}$ clearly reflects the consistency of the
overall analysis adopted here.  In addition, it can be seen clearly
from Fig.~4 that the intrinsic contribution dominates in the whole
temperature range 5 to 300 K.

\begin{figure}
\center
\includegraphics[width=0.42\textwidth]{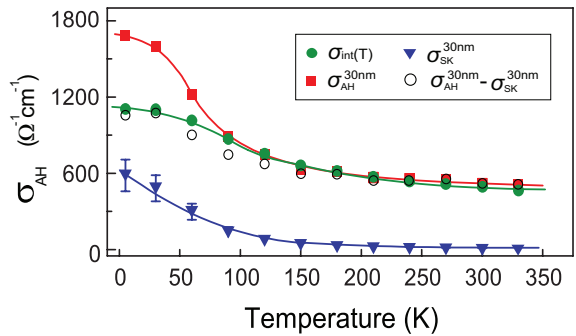}
\caption{ Temperature dependence of $\sigma_\text{int}$ in Ni
together with various contributions in 30nm thick Ni; the solid
curves are guidelines for eyes.}
\end{figure}

The temperature dependence of the intrinsic AHE can be understood
from the existence of degenerate or nearly degenerate bands near the
Fermi surface in Ni.  It is well known that in Ni the minority spin
bands have small hole pockets around $X$\cite{21}.  These bands are
of $t_{2g}$ character and are doubly degenerate in the absence of
the SOC.  With SOC, this degeneracy is lifted, resulting in large
value of the Berry curvature\cite{30}.  Given the close proximity of
these bands to the Fermi surface, they will be thermally populated
at finite temperatures.  First-principles calculations indicate that
contributions to the AHE from these hole pockets are opposite in
sign \cite{14}, which explains why $\sigma_\text{int}(T)$ decreases
with increasing temperature. Similar behavior has also been observed
in first-principles calculation of the spin Hall effect in Pt
\cite{22}.

With our new insight on the AHE in Ni, we now clarify the earlier
confusions and complications in literatures. Given the values of
$\sigma_\text{int}(5 \text{ K})=1100$ $(\Omega\cdot\text{cm})^{-1}$
and $\alpha=-7\times10^{-4}$, the skew scattering term
$\rho_\text{sk}=\alpha\rho_{xx0}$ is expected to overwhelm the
intrinsic one $\rho_\text{int}=-\sigma_\text{int}\rho_{xx}^{2}$ if
the residual resistivity $\rho_{xx0}$ is below 0.5
$\mathrm{\mu\Omega\cdot cm}$, which corresponds to the so called
"clean limit" in the AHE where the linear term
$\rho_{AH0}=\alpha\rho_{xx0}$ dominates. This explains why in the
ultra-pure Ni samples at low temperature, Fert et al., did observe
an overall linear scaling $\rho_\text{AH0}=\alpha\rho_{xx0}$. On the
other hand, the temperature dependent power law scaling
$\rho_{xx}^{n}$ or the average power law $\rho_{xx}^{1.4}$ are in
fact some ill-defined average of the real scaling
$\rho_\text{AH}=\alpha\rho_{xx0}+\beta\rho^{2}_{xx0}-\sigma_\text{int}(T)\rho^{2}_{xx}$.
Finally, so far first-principles calculations of the intrinsic AHE
have been compared to  320 to 750 $(\Omega\cdot\text{cm})^{-1}$
measured at room temperature\cite{2,12}.  Our result shows that for
zero-temperature calculation, one should compare to
$\sigma_\text{int}(5K)=1100$ $(\Omega\cdot\text{cm})^{-1}$.

The authors thank G. Malcolm Stocks for useful discussions on the
band structure of Ni.  This work was supported by MOST (No.\ 2009CB929203),
NSFC (No.\ 10834001), and SCST.  D.X. acknowledges support from the Division of
Materials Sciences and Engineering, Office of Basic Energy Sciences,
U.S. Department of Energy.

\end{document}